\documentclass[12pt,preprint]{aastex}

\shorttitle{Spitzer spectroscopy of OHPNe}
\shortauthors{Garc\'{\i}a-Hern\'andez et al.}

\begin{document}

\title{Spitzer/IRS spectroscopy of high mass precursors to planetary nebulae}


\author{D. A. Garc\'\i a-Hern\'andez\altaffilmark{1}, J. V.
Perea-Calder\'on\altaffilmark{2}, M. Bobrowsky\altaffilmark{3} and P. 
Garc\'\i a-Lario\altaffilmark{4}}

\altaffiltext{1}{The W. J. McDonald Observatory. The University of Texas at
Austin. 1 University Station, C1400. Austin, TX 78712$-$0259, USA; 
agarcia@astro.as.utexas.edu}
\altaffiltext{2}{European Space Astronomy Centre, INSA S. A., P.O. Box  50727. E-28080 Madrid. Spain;
Jose.Perea@sciops.esa.int}
\altaffiltext{3}{Computer Sciences Corporation/Space Telescope Science 
Institute. Baltimore MD, USA;
matt@mailaps.org}
\altaffiltext{4}{Herschel Science Centre. European Space Astronomy Centre,
Research and
Scientific Support Department of ESA. Villafranca del Castillo, P.O. Box
50727. E-28080 Madrid. Spain; 
Pedro.Garcia-Lario@sciops.esa.int}

\begin{abstract}
We present Spitzer/IRS observations of a small sample of heavily obscured
IRAS sources displaying both the infrared and OH maser emission characteristic
of OH/IR stars on the asymptotic giant branch (AGB), but also radio continuum
emission typical of ionized planetary nebulae (PNe), the so-called {\it
OHPNe}. Our observations show that their mid-infrared spectra are dominated by
the simultaneous presence of strong and broad amorphous silicate absorption
features together with crystalline silicate features, originated in their O-rich
circumstellar shells. Out of the five sources observed, three of them are
clearly non-variable at infrared wavelengths, confirming their post-AGB status,
while the remaining two still show strong photometric fluctuations, and may
still have not yet departed from the AGB. One of the non-variable sources in the
sample, IRAS 17393$-$2727, displays a strong [Ne II] nebular emission at 12.8 
$\mu$m, indicating that the ionization of its central region has already
started.  This suggests a rapid evolution from the AGB to the PN stage. We
propose that these heavily obscured OHPNe represent the population of high mass
precursors to PNe in our Galaxy.
\end{abstract}


\section{Introduction}
 The evolution of low- and intermediate-mass stars (1$-$8 M$_{\odot}$)  ends
with a strong mass loss phase at the tip of the asymptotic giant branch  (AGB)
and the subsequent formation of a new planetary nebula (PN) (e.g. Kwok
2000). The transition time required for an AGB star to become a  PN (the
so-called  ``post-AGB" phase) depends essentially on the initial mass of the
progenitor star, and it ranges from just a few hundred years for  the most
massive stars, to several thousand years for low mass progenitors (e.g.
Bl\"ocker 1995).  Crucial changes affecting both the morphology and  the
chemistry of the central star also occur during this short-lived  post-AGB phase
which completely determine the subsequent evolution of the  star as a PN (see
e.g. van Winckel 2003).

Most candidate post-AGB stars were identified as such now more than 20
years ago, based on the infrared photometry provided by IRAS, as these stars
show characteristic infrared colours (e.g. Garc\'\i a-Lario et al. 1997, and
references therein). However, the confirmation of their post-AGB nature has only
been systematically explored more recently, through  extensive ground-based
follow-up spectroscopic surveys in the optical (Garc\'\i a-Lario 2006, and
references therein). As a consequence of this, a few hundred candidates have
been identified as genuine post-AGB stars (e.g. Su\'arez et al. 2006).
However, it is suspected that a strong bias exists in the type of stars
identified this way since they may only represent the slowly evolving,
low-mass population of post-AGB stars in the Galaxy. These low-mass stars
experience lower mass loss rates at the end of the AGB phase and thus are more
easily detectable in the optical as bright stars of intermediate spectral type.
In contrast, very little is known about the population of more massive, heavily
obscured post-AGB stars which may have systematically escaped detection in the
classical optical surveys carried out so far. These stars may only be detectable
at infrared wavelengths, and they could be the actual precursors of most PNe in
our Galaxy. 

 Among this hidden post-AGB population, Zijlstra et al. (1989) identified a 
peculiar, very small group of heavily obscured IRAS sources showing 
both the characteristics of standard OH/IR stars (a strong infrared emission 
accompanied by the detection of the OH maser emission at 1612 MHz) and PNe
(radio continuum emission at centimeter wavelengths). These sources, called for
this reason {\it OHPNe}, were suggested to represent the `evolutionary
link' between OH/IR stars and PNe. In most cases, however, the evolutionary
stage of the central star  was never confirmed observationally, mainly because
of the high extinction affecting the whole nebula, which makes the obtention of
a useful spectrum of the central star very difficult. This task is further
complicated by the uncertainty in the association between the infrared (IRAS)
and the radio continuum source, a consequence of the poor astrometric quality of
the IRAS Point Source Catalog (PSC) (as large as 1 arcmin in highly
confused, crowded areas close to the Galactic Center, where most of  these
sources are located). 

 With the advent of other, more recent infrared missions like MSX (Egan  et al.
2003) and 2MASS (Cutri et al. 2003), the situation has significantly  improved
and the infrared positions of these OHPN stars can now be determined  to an
accuracy of a few arcsec in the case of MSX, and to the subarcsec level if they
are detected by 2MASS. In addition, with ISO/SWS and Spitzer/IRS spectroscopy we
can now study the properties of the dust  grains formed in their circumstellar
shells. In this letter we present Spitzer/IRS spectra ($\sim$5$-$37 $\mu$m)
obtained on a representative sample of galactic `OHPNe' in order to
elucidate their nature and evolutionary stage.

\section{Observations and data reduction}
A sample of 5 OHPNe: namely, IRAS 17168$-$3736, IRAS 17393$-$2727, IRAS
17418$-$2713, IRAS 17443$-$2949 and IRAS 17580$-$3111 (hereafter I17168,
I17393, I17418, I17443 and I17580, respectively), all them extremely bright
sources at mid-infrared wavelengths (see Table 1) was selected from the
`GLMP catalog of IRAS sources with infrared colors similar to those of known
PNe' (Garc\'\i a-Lario 1992) and observed by us recently with the Infrared
Spectrograph (IRS, Houck et al. 2004) on board the Spitzer Space Telescope
(Werner et al. 2004). All of them have been identified as stars showing
both OH maser emission at 1612 MHz and radio continuum emission (see Table 2)
and thus classified as OHPNe (three of them were indeed already included in the
original paper by Zijlstra et al. 1989). The observations were carried out
between March and April 2005, in the 9.9$-$37.2 $\mu$m range, using the
Short-High (SH) and Long-High (LH) modules (R$\sim$600). Two of the sources in
the sample (I17393 and I17580) were also observed at lower resolution with the
Short-Low (64$<$R$<$128 SL; 5.2$-$14.5 $\mu$m) module. Note that for such
bright stars it is not possible to perform target acquisition using
on-source peak-up (neither peak-up on a nearby source helps) and our astrometric
accuracy is  actually limited by the MSX accuracy, of the order of 1.8", which
is of the order of (or even  worse than) the nominal pointing accuracy of
Spitzer ($\sim$1"). This in principle should be enough to avoid photometric
losses when centering the source in the Spitzer apertures, at least at the
longer wavelengths. In all cases, it was easy to achieve a S/N larger than 50
over the whole spectral range with just 2 exposure cycles of  6 s in each of the
three modules SL, SH, and LH. We started our analysis from the co-added 2-D
flat-fielded images (one for each nod position; pipeline version 12.3)
which were cleaned from bad pixels. The spectra for each nod position were
extracted from the 2-D images, wavelength and flux calibrated using the Spitzer
IRS Custom Extractor (SPICE) with a point source aperture. Since no sky
(off-position) measurements were taken, the contribution of the expected 
background flux through each slit was estimated from the model of Reach and
coworkers\footnote{See http://ssc.spitzer.caltech.edu/documents/background/} and
subtracted from the SH and LH spectra. For the module SL the two nod positions
were previously differenced in order to subtract the sky background. The 1-D
spectra were cleaned for residual bad pixels, spurious jumps and glitches,
smoothed and merged into one final spectrum per module for each source using the
Spitzer contributed software SMART (Higdon et al. 2004). 

In general, we found a very good match between the different nod position 
spectra except at wavelengths longer than 34 $\mu$m. This spectral range
corresponds to the red end (order 11) of the LH module, affected by a strong
noise level,  and was for this reason excluded from our analysis.  In addition,
we found a flux deficit of $\sim$20 \% in the  SL module spectrum of I17580 with
respect to the SH observations of the same source. We attribute this mismatch to
the slightly extended nature of  I17580 at infrared wavelengths and the
different apertures used by these observing modes. This interpretation is also
consistent with the extended nature (6.7$\arcsec$) of the associated radio
continuum emission (see Table 2). At longer wavelengths, however, the absolute
flux level of the LH module spectra was found to be $\sim$10$-$15 \% higher for
all the sources in our sample  with respect to the corresponding SH module
observations. In principle, this effect would be expected in case of the sources
being extended. However, we suspect that a systematic effect is affecting our LH
data. This is because for I17393 (the other source observed in the SL module)
the match between the SL and SH modules is perfect, confirming its point source
nature (see Table 2) while the LH observations show a similar flux excess.
In addition, the non-variable sources in our sample (I17168, I17393 and I17580;
see Sect. 4), show as well LH fluxes which are systematically higher than
the MSX and the IRAS ones at this wavelength range while the SH fluxes are in
perfect agreement with the photometric data.  Thus, we scaled the LH spectra to
the SH spectra in order to obtain one final spectrum per source. 

\section{Results}
The 5 OHPNe are listed again in Table 2 together with the 
some additional information available for them in the literature such as the
main characteristics of their OH maser and radio continuum emission, or
the IRAS variability index, as quoted in the IRAS PSC (Beichman
et al. 1988). OHPN stars in our sample are so heavily obscured that they are
completely invisible in the optical domain, even at wavelengths below 3 $\mu$m.
I17580 is the only source in our sample for which a near-IR counterpart (K=9.3)
was found in the 2MASS catalog. During the observations, Spitzer was
pointed to the improved coordinates (with respect to IRAS) quoted in the MSX
PSC, except for I17580, for which we used the astrometric
information in the 2MASS PSC, accurate to within the subarcsec
level. These coordinates are compared in Table 2 with the available OH and radio
continuum positions. The OH positions nicely coincide within the errors with
the Spitzer ones in all cases, confirming that the OH maser emission is
associated with the infrared source. Unfortunately, radio continuum
measurements date from more than 10 years ago and in some cases it is not
possible to make a proper analysis of the associated astrometric uncertainties,
although at least for I17168, I17393 and I17418 the radio continuum positions
nicely coincide with the OH and infrared positions within a few arcsec.

The reduced Spitzer/IRS spectra are shown in Fig.1. The O-rich chemistry is
confirmed by the simultaneous detection of strong and broad amorphous silicate
absorptions at 9.7 and 18 $\mu$m (the latter less obvious in I17580 and I17393)
in all sources, together with complex crystalline silicate profiles generally
attributed to olivine and pyroxenes with various mixtures of Mg and Fe (e.g.
those at $\sim$13$-$16, $\sim$19$-$21, 23$-$25, 26$-$28, and 29$-$31 $\mu$m; see
Koike et al. 2000). The features are seen either in absorption or in emission  
with a variety of strengths and at slightly different wavelengths from source to
 source, depending on the specific chemical composition and size of
the dust grains in the circumstellar envelope. In
particular, we detect crystalline silicate features in emission at 27.8, 29.3,
and 30.5 $\mu$m (see Fig.1). The 27.8 $\mu$m feature can be attributed to
forsterite (a Mg-rich crystalline silicate), while the features around 29.3 and
30.5 $\mu$m remain unidentified (Molster et al. 2002). At shorter wavelengths,
some of these crystalline features appear in absorption or affected by
self-absorption (e.g. those features centred at $\sim$13.6, 14.4, 15.4, 19.5,
20.6, 23.6, and 26.6 $\mu$m), indicating the large optical depth of the
circumstellar shell. The 13.6 and 14.4 $\mu$m features are clearly detected in I17443,
less evident in I17418 and I17168 and not seen in I17580 and I17393. The
detection of crystalline silicates in absorption has previously been reported in
some other extreme OH/IR stars observed with ISO, such as OH 26.5$+$0.6
(Sylvester et al. 1999) also shown in Fig.1, for comparison. In addition, other
unidentified crystalline silicates may also be present in the spectra of our
stars, suggesting that other, more unusual dust species may be present
in their circumstellar shells.

We have classified the stars in our sample according to the decreasing strength
of the amorphous silicate absorption features at 9.7 and 18 $\mu$m (from top to
bottom in Fig.1) in what  it seems to be an evolutionary sequence. The
first three sources in Fig.1; I17443, I17418 and I17168, show spectra which are
very similar to those observed in other extreme OH/IR stars (Sylvester et
al. 1999), and are dominated by the strong and broad amorphous silicate
absorption features at 9.7 and 18 $\mu$m, with emerging crystalline silicate
emission features which are more prominent at longer wavelengths.
The two other sources in Fig.1 I17580 and I17393, are probably in a more
evolved stage and show much fainter amorphous silicate absorption features at
9.7 and 18 $\mu$m. These sources are peculiar in the sense that they display
extremely red colors, indicative of a cool expanding circumstellar shell.
I17580 seems to be in a less evolved stage compared to I17393, as
suggested by the flux drop observed at 60 $\mu$m by IRAS. In contrast, I17393
seems to be more evolved, with cooler dust in the shell as deduced from
the IRAS photometry at 60 $\mu$m and, more interesting, with a strong [Ne II]
nebular emission at 12.8 $\mu$m (see Fig.1), suggesting that the onset of
ionization has already taken place. 

\section{Variability analysis}

Interestingly, we found a good consistency between the Spitzer/IRS spectra and
the photometric flux densities measured at different epochs by IRAS and MSX for
I17168, I17393 and I17580, confirming their status of non-variable
sources, as expected from their low IRAS variability index (VAR column in Table
2). However, this is not the case of I17418 and I17443, for which the IRAS
variability index is indeed higher. These sources seem to be strongly variable
and may still be evolving as AGB stars (see Sect. 5). The flux differences
between our final Spitzer/IRS spectra (after the correction of the
10$-$15 \% flux excess observed in the LH module) and the large photometric
variations observed at different epochs (more than 30 \% in flux at 25 $\mu$m),
cannot be attributed to color-corrections because in the range of temperatures
of the sources in our sample ($\sim$100$-$300 K) they are always lower than 5
and 15 \% for the MSX and IRAS photometry, respectively. A possible
contamination of the IRAS fluxes by nearby infrared sources can also be
discarded, according to the available information in the IRAS and MSX Point
Source Catalogs. The variability status of these two sources is also confirmed
by the different flux levels measured in the ISO PHOT-S spectra with respect to
our Spitzer spectra at the overlapped wavelength regions (see Fig.1).

\section{OHPNe: the high-mass precursors to PNe?}

The stars in our sample belong to a small group of O-rich AGB stars 
characterised by their strong obscuration in the optical, which have  recently
been suggested to represent the missing population of massive AGB stars in our
Galaxy. These are extreme OH/IR stars showing the larger OH expansion velocities
(V$_{exp}(OH)$ $>$ 12 km s$^{-1}$; see Table 2) and the thicker circumstellar
envelopes, being usually invisible in the optical domain, as it is the case of
the 5 sources here studied. With ISO only a few other obscured massive O-rich
AGB stars were observed. Those sources with available ISO spectroscopy  display
also strong absorption bands of amorphous silicates together with crystalline
silicate features (see e.g. Sylvester et al. 1999), resembling the OHPNe
presented in this paper. Analyzing ISO-SWS spectra, Garc\'\i a-Lario \&
Perea-Calder\'on (2003) found that dramatic changes occur in the infrared SEDs
of these stars during this phase of total obscuration. In the case of massive
O-rich AGB stars the strong silicate absorption features at 9.7 and 18 $\mu$m
are seen first in combination with prominent crystalline silicate features which
emerge covering the range from 10 to 45 $\mu$m. Later, these features become
dominant and are sometimes still observed in more evolved  O-rich PNe (Molster
et al. 2001). It has been suggested that high temperature crystallization can
take place at the very end of the AGB phase as a consequence of the strong mass
loss  (Waters et al. 1996; Sylvester et al. 1999). Alternatively, low
temperature crystallization is also predicted in long-lived circumbinary disks
surrounding, leading to a different pattern of crystalline silicates (Molster et
al. 1999) not observed in our sources. It should be noted that OHPNe in our
sample (specially the non-variable ones) may be already developing strong
bipolar post-AGB outflows, as it is also  observed in bipolar Type I PNe, and it
is likely that  a thick circumstellar disk/torus (where the crystallization
could take place) is surrounding the central source. Our Spitzer spectra reveals
with an unprecedented level of detail and better temporal resolution than
previous ISO observations the evolution of the dust features at the precise
moment when the transition from the AGB to the post-AGB stage is taking place.
We basically observe how the strong amorphous silicate absorption features which
are characteristic of AGB stars (see the spectra of I17418 and I17443) disappear
in very short timescales (see below) leading to a totally different spectrum
dominated by crystalline silicate features (seen either in absorption or in
emission depending on the optical depth of the shell at different wavelengths),
until the gas in the circumstellar envelope gets ionized as a consequence of the
rapid increase of the effective temperature of the central star (as it is the
case of I17393).

It is well known that the more massive AGB stars (M $\gtrsim$ 4$-$5 M$_\odot$ at
solar metallicity) can evolve all the way from the AGB phase to the PN stage as
O-rich stars as a consequence of the activation of the so-called ``hot bottom
burning (HBB)" process (e.g. Mazzitelli et al. 1999), which prevents the
formation of C-rich AGB stars. Recently, some of the sources displaying  a
faint, but still visible optical counterpart, have been observed by us
(Garc\'\i a-Hern\'andez et al. 2006, 2007) and strong overabundances of lithium and
rubidium have been found, confirming their status of HBB AGB stars. If our
sources are indeed massive HBB AGB stars, as suggested by the O-rich chemical
composition, the strong obscuration and the high expansion velocity observed, we
would expect a very fast evolution to the PN stage and thus a young dynamical
age for the  dust in the envelope. If we perform a simple  calculation for
I17393 (the most evolved object in our sample which is already surrounded by
ionized  gas), we found that the short wavelength  range of the SED (from 5 to
16 $\mu$m) can be well reproduced by  dust emitting at 166 K. By assuming a
star's luminosity of 10,000 L$_\odot$ (representative of stars at the tip of the
AGB phase), the equilibrium radius of astronomical silicates with a typical
grain size of 0.1 $\mu$m is found to be $\sim$0.003 pc. The expansion velocity
(14.6 km s$^{-1}$ as derived from the OH maser emission) give us a dynamical age
of the circumstellar envelope of I17393 of only $\sim$205 years. This is
consistent with the expected transition time for massive post-AGB stars
($\sim$100$-$1000 yr for M $\gtrsim$ 4 M$_\odot$; e.g. Bl\"ocker 1995), which
confirms our hypothesis.

In summary, our observations show that at least three of the observed OHPNe are
massive non-variable O-rich stars which have left the AGB very recently. They
must be rapidly evolving towards the PN stage representing the evolutionary link
between massive OH/IR stars and PNe. The detection of [Ne II] nebular emission
in the heavily obscured OHPN star I17393 provides the first observational
evidence that the evolution of some of these massive post-AGB stars can be so
fast ($\sim$100$-$1000 yr for M $\gtrsim$ 4 M$_\odot$; e.g. Bl\"ocker 1995) that
the temperature of the central star may be hot enough to ionize the gas in the
circumstellar envelope before it gets diluted into the interstellar medium. For
this to happen, the ionization must start within a few hundred years after the
AGB mass loss ends. Otherwise, the envelope will have dispersed and the OH
emission stopped as it cannot be sustained for too long once the mass loss rate
drops by several orders of magnitude at the end of the AGB phase. However, we
have also shown here that two of the observed OHPNe are strongly variable O-rich
stars and they may still be evolving as AGB stars. At present, the simultaneous
detection of radio continuum emission from these sources is not well understood,
unless this type of emission has a strong non-thermal contribution which arises
from shock interactions at the interface between a recently developed fast wind
and previously ejected material. This is supported by the recent detection of
strong non-thermal radio continuum emission in the massive (M $\gtrsim$ 4
M$_\odot$) O-rich AGB star V1018 Sco (Cohen et al. 2006). We propose that these
OHPNe represent the high mass precursors to PNe in the Galaxy.

\acknowledgments
This work is based on observations made with the Spitzer Space Telescope, 
which is operated by the Jet Propulsion Laboratory, California Institute of
Technology, under NASA contract 1407. JVPC and PGL acknowledge support 
from grant \emph{AYA 2003$-$9499 from the Spanish Ministerio de Educaci\'on y
Ciencia}. MB acknowledges support from Grant 3633 from the Spitzer Space
Telescope.

\clearpage

\begin{deluxetable}{lccccccc}
\tabletypesize{\scriptsize}
\tablecaption{MSX and IRAS photometry \label{tbl-1}}
\tablewidth{0pt}
\tablehead{
\colhead{IRAS Name} &   \colhead{MSX$_{A}$} & \colhead{MSX$_{C}$} &
\colhead{MSX$_{D}$} &
\colhead{MSX$_{E}$} & \colhead{IRAS$_{12}$} &\colhead{IRAS$_{25}$} 
&\colhead{IRAS$_{60}$}\\
\colhead{} &   \colhead{(Jy)} & \colhead{(Jy)} &
\colhead{(Jy)} &
\colhead{(Jy)} & \colhead{(Jy)} &\colhead{(Jy)} 
&\colhead{(Jy)}
}
\startdata
17168$-$3736 & 1.79 & 10.12 & 23.05 & 24.95 & 9.78 & 37.00& 47.90 \\ 
17393$-$2727 & 0.25 &  1.05 &  2.97 & 10.72 & 1.83 & 17.80& 36.80 \\ 
17418$-$2713 & 17.92 & 29.79 & 58.00 & 68.32 &14.90 & 51.30& 45.90 \\
17443$-$2949 &15.11 & 22.46 & 42.91 & 39.45 &15.80 & 39.40& 34.50 \\
17580$-$3111 & 1.24 &  3.62 &  6.87 & 13.16 & 3.23 & 15.30&  7.96 \\
\enddata
\end{deluxetable}


\begin{deluxetable}{lccccc}
\tabletypesize{\scriptsize}
\tablecaption{The sample of OHPNe and astrometric information$^{a}$.\label{tbl-2}}
\tablewidth{0pt}
\tablehead{
\colhead{Object} & \colhead{V$_{exp}$(OH)$^{b}$} &
\colhead{(F$_{6cm}$, Diam.)$^{c}$} & \colhead{Var} & \colhead{($\Delta\alpha$, $\Delta\delta$)}& \colhead{($\Delta\alpha$, $\Delta\delta$)}\\
\colhead{} & \colhead{(km s$^{-1}$)} & \colhead{(mJy, $''$)} & \colhead{} &
\colhead{Spitzer-OH($''$)} &  \colhead{Spitzer-Radio($''$)}\\
}
\startdata
17168$^{e}$       & 13.1 (1)   &(54$^{d}$, $<$1) (1) &28 & ($+$0.75, $+$0.70) &($+$0.45, $+$3.5)  \\
17393 & 14.6 (1)   & (1.4, $<$1) (2)     &18 & ($-$0.90, $+$0.40) &($+$0.15, $-$0.10) \\
17418       & 15.3 (1)   & (2.0, $<$1) (3)     &99 & ($+$1.20, $-$0.40) &($-$4.35,$-$2.50) \\
17443       &$\dots$ (2) & (0.9, 3.4) (4)      &51 & ($+$0.00, $-$3.00) &($-$7.50, $+$6.80) \\
17580$^{f}$ &14 (3)      & (2.5, 6.7) (4)      &9  & ($+$12.0, $+$0.00) &($-$12.00, $-$30.2) \\
\enddata

\tablenotetext{a}{The Spitzer coordinates are taken either from the MSX or the
2MASS Point Source Catalogs (see text ).} 
\tablenotetext{b}{References for OH  data: (1) Sevenster et al. 1997; (2) David
et al. (1993); (3) Hu et al. (1994)}
\tablenotetext{c}{References for radio continuum observations: (1)
Zoonematkermani et al. (1990), (2) Pottasch et al. (1987), (3) Ratag 1990,
(4) Ratag et al.(1990)}
\tablenotetext{d}{Radio continuum flux at 20 cm (Zoonematkermani et al. 1990)}
\tablenotetext{e}{Improved radio coordinates were taken from the new version of
the MAGPIS catalog (http://third.ucllnl.org/gps/), Robert H. Becker (private
communication)}
\tablenotetext{f}{Ziljstra et al. (1989) and Hu et al. (1994) reported a coincidence
of the OH position with the infrared one}
\end{deluxetable}

\clearpage

\begin{figure}
\epsscale{0.7}
\plotone{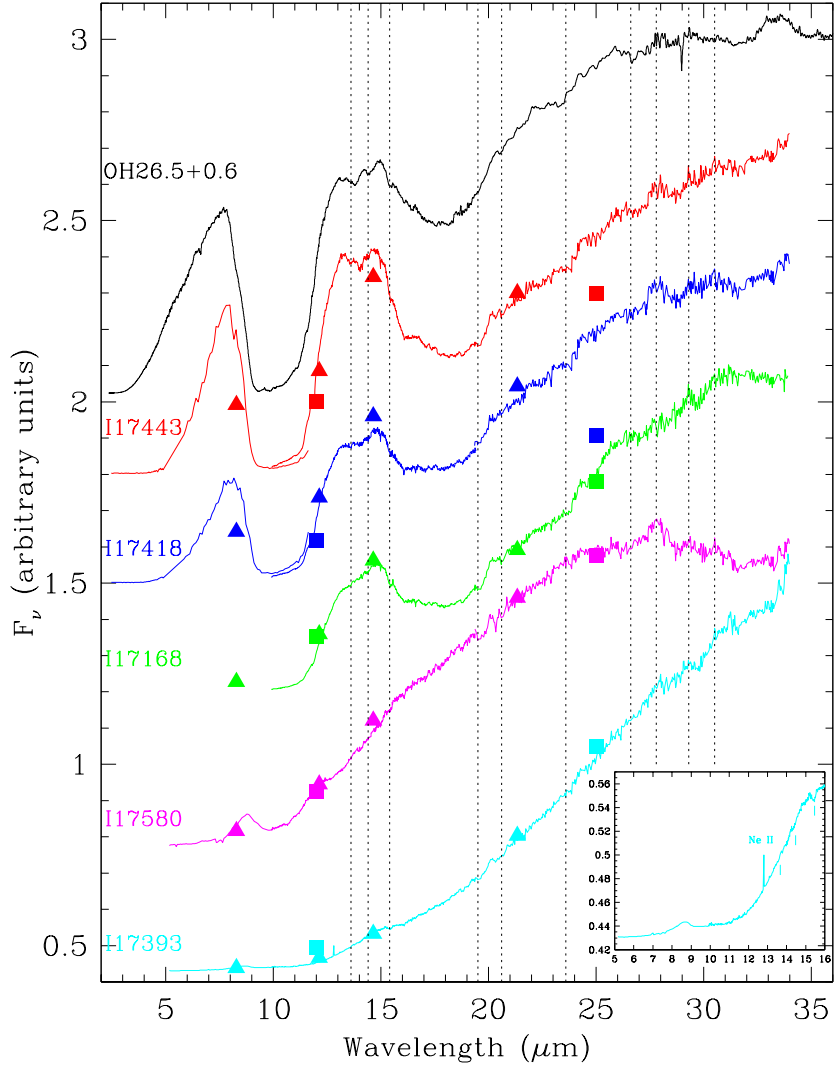}
\caption{The Spitzer/IRS spectra of the 5 OHPN stars observed are shown in
comparison to the ISO spectrum of the extreme OH/IR star OH 26.5$+$0.6 (at the
top). MSX (triangles) and IRAS (squares) photometry is also displayed for our
target stars. The spectra are shown (from top to bottom) according the
decreasing strength of the amorphous silicate absorption features at 9.7 and 18
$\mu$m. The theoretical positions of those crystalline silicate features
which appear either in absorption or in emission  mentioned in the text are
indicated with dashed vertical lines. Note that for I17443 and I17418, the
PHOT-S spectra (R$\sim$90; 2.5$-$11.6 $\mu$m) available in the ISO Data Archive
are also displayed. \label{fig1}}
\end{figure}

\end{document}